\documentclass[aps,prl,twocolumn,groupedaddress]{revtex4}

\usepackage{graphicx}  

\begin{document}


\title{Control of Raman lasing in the non-impulsive regime}


\author{B. J. Pearson}
\author{P. H. Bucksbaum}
\affiliation{FOCUS Center and Physics Department, University of
Michigan, Ann Arbor, MI 48109-1120}


\date{\today}

\begin{abstract}
We explore coherent control of stimulated Raman scattering in the
non-impulsive regime. Optical pulse shaping of the coherent pump
field leads to control over the stimulated Raman output. A model
of the control mechanism is investigated.
\end{abstract}


\maketitle


Since the advent of pulsed lasers, there has been extensive
experimental and theoretical work on stimulated Raman scattering.
More recently there has been a resurgence of interest in impulsive
Raman
scattering\cite{nelson:dec1985,nelson:mar1990,korn:jul1998,weinacht:jan2002b}
due in part to the development of the field of learning coherent
control\cite{rabitz:mar1992,warren:mar1993,rabitz:may2000}, along
with advances in ultrafast laser technology\cite{backus:mar1998}
and programmable pulse shaping\cite{nelson:may1995,
warren:jan1997}. In the impulsive regime, the laser bandwidth is
large compared to the frequency of the Raman mode (typically a
molecular vibration). In the time domain the pulse duration is
short compared to the vibrational period, and one can shape the
optical pulse to drive molecular mode(s) on
resonance\cite{weinacht:aug2001}. Alternatively, if the laser
bandwidth contains photon pairs separated by the Stokes frequency,
then the Raman gain can be seeded by the pump laser. Control of
the Raman gain can be achieved by appropriately phasing colors in
the pump pulse.

For the case of two-photon atomic absorption, this control
mechanism has been described as shaping of the nonlinear power
spectrum of the driving
field\cite{silberberg:nov1998,bucksbaum:nov1998,silberberg:aug1999}.
The idea has since been extended to multi-photon absorption in
molecules \cite{dantus:oct2002,dantus:feb2003}, as well as
vibrational Raman excitation in multi-mode molecular
systems\cite{weinacht:aug2001}. Here again analysis leads to an
explanation of the control via the nonlinear power spectrum of the
pump pulse.

In the non-impulsive regime, the laser bandwidth is small compared
to the frequency of the mode. Since there are no photon pairs
separated by Stokes frequency in the pump pulse, the Stokes field
must build up from spontaneous Raman scattering.  This produces a
Stokes field whose phase is random and cannot be controlled.
Control may still be possible over the stimulated output spectrum
or the final state populations in the presence of multiple Raman
modes. Non-impulsive control of the stimulated Raman spectrum in
liquid methanol has been reported in experiments that used
learning algorithms to discover the optimal driving field
\cite{pearson:may2001,pearson:jan2002}. Here we demonstrate a
possible mechanism for this control.

The details of our adaptive learning technique and laser system
have been described previously \cite{pearson:may2001}. In learning
control experiments, the physical system is used to find the
optimal driving field (pulse shapes) without prior knowledge of
the Hamiltonian \cite{rabitz:mar1992}.  These solutions can be
analyzed later to learn about the underlying quantum dynamics.
Briefly, our experiments use a shaped, ultrafast Ti:Sapphire laser
system as the excitation source. The laser pulses are shaped in an
acousto-optic Fourier filter \cite{warren:jan1997} interfaced with
a computer. Our spectral bandwidth ($4-5$ THz) and pulse shaping
characteristics provide temporal control over the pulses ranging
from $100$ fs to $5$ ps in duration. Our adaptive learning
algorithm determines the pulse shapes by optimizing a feedback
signal derived from the Raman spectrum.

\begin{figure}[ht]
\includegraphics [width=8.5cm, height=8.5cm, angle=0] {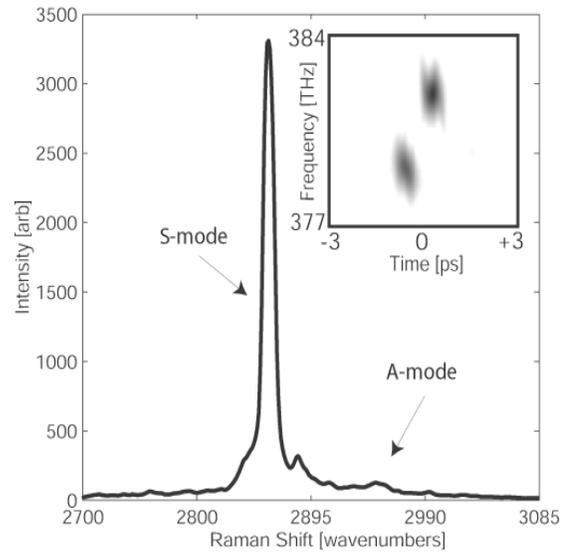}
\caption{Resulting Raman spectra obtained after optimization with
the feedback algorithm. Inset shows a Husimi plot of the optimal
``double-blob" pulse shape that led to S-mode excitation.}
\label{2blobexp:husspec}
\end{figure}
The system under investigation in these experiments is liquid
phase methanol (CH$_{3}$OH). We focus on the two C-H stretch
vibrational modes labelled S ($2834$ cm$^{-1}$) and A ($2946$
cm$^{-1}$) \cite{shim:1972}. The experiments are performed in a
$10$ cm liquid cell, and the forward scattered Raman spectrum is
collected and fed back to the algorithm.  The learning algorithm
is able to channel the gain into either of the two Raman modes.
Although the algorithm finds a variety of pulse shapes that
produce the desired effect, one class of solutions stands out.
Figure \ref{2blobexp:husspec} (inset) shows a representative pulse
shape of this class of solutions in the form of an optical Husimi
distribution\cite{wigner:apr1984}.  The Husimi plot is a
two-dimension convolution of the Wigner
function\cite{wigner:apr1984,cohen:jul1989,paye:oct1992} of the
pulse shape that approximates the intensity distribution in both
time and frequency. The optimal pulse shapes found by the
algorithm consist of a pair of nearly transform-limited ``blobs"
that are separated by approximately $3.3$ THz. This energy
splitting matches the S-A mode spacing. Figure
\ref{2blobexp:husspec} shows the resulting Raman spectrum after
excitation by the pulse shown in the inset of Fig.
\ref{2blobexp:husspec}. The pulse produces strong selective
excitation of the S vibrational mode.

\begin{figure}[ht]
\includegraphics [width=8.5cm, height=5.7cm, angle=0] {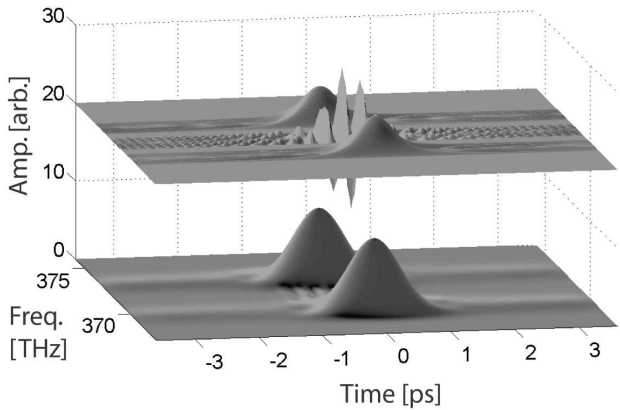}
\caption{Calculated Wigner plots for double-blob pulse used in the
experiment when the phase offset between the blobs is equal to
$\pi$. The lower plot is the Husimi distribution when the phase
offset is zero.} \label{2blob:huswig}
\end{figure}

These two-blob solutions suggest a quasi-impulsive model for the
interaction, where the pump field directs the Raman gain through a
coherent coupling between the two modes \cite{pearson:may2001}.
This coherence modulates the intensity envelope of the pulse.  The
modulation is apparent in the Wigner distribution, but has only a
minimal effect on the Husimi distribution (Fig. $2$).

In order to study the effect of the phase offset, we excite the
molecules with the original pulse shape shown in Fig.
\ref{2blob:huswig}. We then collect the stimulated Raman spectra
as the phase offset between the two blobs is varied from $0$ to
$\pi$.
\begin{figure}[ht]
\includegraphics [width=8.5cm, height=8.5cm, angle=0] {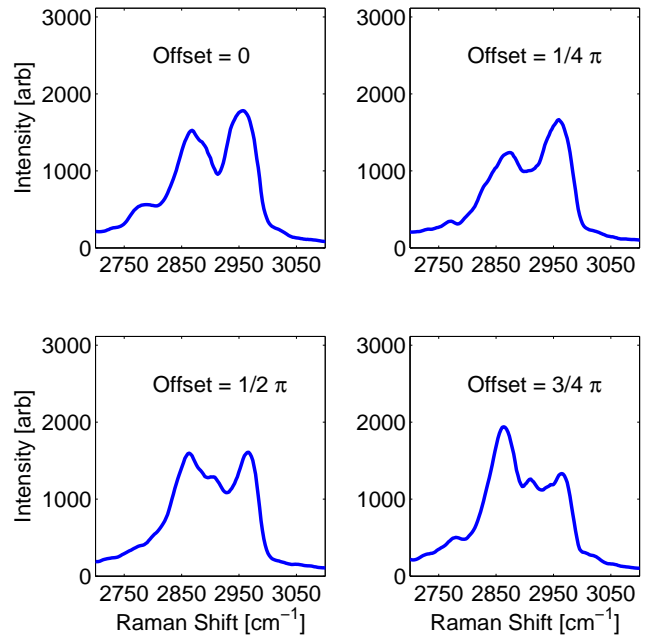}
\caption{Stimulated Stokes output as a function of the Raman shift
from the central frequency of the pump pulse.  The different
panels show the output spectra for various phase offsets. The gain
oscillates between the S- and A-modes.} \label{phasescan:2d}
\end{figure}
Figure \ref{phasescan:2d} shows the stimulated Raman output
spectrum for several different phase offsets. As the offset is
varied, the Raman gain oscillates between the two different Stokes
modes.  As is to be expected, this single parameter control is not
as effective at mode selection as the full learning algorithm (see
Fig. \ref{2blobexp:husspec}).  Nevertheless, control is achieved
by varying only the phase offset. Care was taken to adjust the
laser intensity to avoid saturation of the Raman gain, since this
leads to significant changes in the observed control.

A simple model that predicts the observed effect can be
constructed by expanding on previous work on single mode Raman
scattering. The theory governing stimulated Raman gain of a single
active mode under excitation by an off-resonant pump pulse has
been developed both semi-classically \cite{bloembergen:jul1970}
and quantum mechanically \cite{raymer:oct1981,raymer:jan1990}. In
the quantum case, the Stokes field in the non-impulsive regime is
generated by spontaneous fluctuations of the Raman polarizability.
A number of approximations can be made to reveal the underlying
physics.  In the standard treatment, one assumes one-dimension,
plane-wave pulse propagation in the slowly-varying envelope
approximation (SVEA), and the calculation is performed in the
transient regime where damping can be neglected. The pulse
durations ($0.1$ to $1$ ps) imply that the SVEA is valid. The
assumption of transient scattering may be an oversimplification
because the pulse lengths approach the decoherence times of the
C-H vibrations\cite{dlott:may2000}. With these simplifications,
the interaction can be reduced to a set of coupled differential
equations describing the propagation along $\hat{z}$ of three
fields inside the medium \cite{lewenstein:mar1984}:
\begin{eqnarray}
\frac{\partial}{\partial x}\varepsilon_{L}=-q\varepsilon_{S}
\label{eq:scaled-pump},
\\
\frac{\partial}{\partial x}\varepsilon_{S}=q^{*}\varepsilon_{L}
\label{eq:scaled-stokes},
\\
\frac{\partial}{\partial
\tau}q=\frac{1}{4}\varepsilon_{L}\varepsilon_{S}^{*}
\label{eq:scaled-q}.
\end{eqnarray}
In these equations, $\varepsilon_{L}$ is the pump laser field,
$\varepsilon_{S}$ is the Stokes field, and $q$ is the molecular
polarizability.  The independent variables $x \propto z$ and $\tau
\propto (t-z/c)$ are the reduced space-time coordinates.

\begin{figure}[ht]
\includegraphics [width=6cm, height=6cm, angle=0] {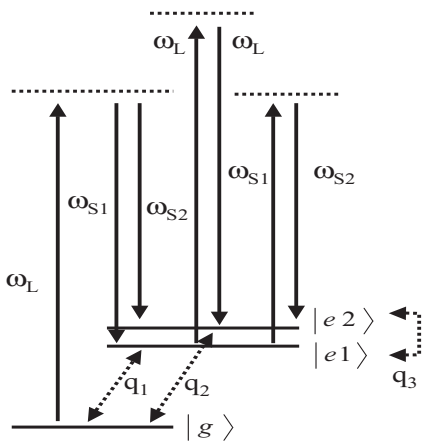}
\caption{Energy level diagram with two Raman-active vibrational
modes.  The three levels are Raman coupled by the pump
($\omega_{L}$) and Stokes ($\omega_{S1}$ and $\omega_{S2}$)
fields.  The interaction generates coherences between pairs of the
three levels ($q_{1}$, $q_{2}$, and $q_{3}$).  An additional
interaction is included by allowing a two-photon Raman coupling
between the two excited vibrational states driven by either the
applied pump field or the generated Stokes fields.}
\label{energylevels}
\end{figure}

We have extended this treatment to include two vibrational Raman
modes which, in addition to the original coupling to the same
ground state, are Raman coupled to each other through two-photon
transitions driven by the applied pump field or the generated
Stokes fields. The relevant energy levels are shown in Fig.
\ref{energylevels}, where $|e1\rangle$ and $|e2\rangle$ are the
first excited levels of the two vibrational modes.  Equations
\ref{eq:scaled-full-pump} through \ref{eq:scaled-q3} are the
modified equations governing the optical fields and molecular
polarizabilities in the case of two modes,
\begin{eqnarray}
\frac{\partial}{\partial x}\varepsilon_{L} &=& - \varepsilon_{S1}
q_{1} -  \varepsilon_{S2} q_{2} - \frac{1}{4} \varepsilon_{L}
q_{3} + \frac{1}{4} \varepsilon_{L} q_{3}^{*}
\label{eq:scaled-full-pump},
\\
\frac{\partial}{\partial x}\varepsilon_{S1} &=& \varepsilon_{L}
q_{1}^{*} + \frac{1}{4} \varepsilon_{S2} q_{3}^{*}
\label{eq:scaled-full-stokes1},
\\
\frac{\partial}{\partial x}\varepsilon_{S2} &=& \varepsilon_{L}
q_{2}^{*} - \frac{1}{4} \varepsilon_{S1} q_{3}
\label{eq:scaled-full-stokes2},
\\
\frac{\partial}{\partial \tau}q_{1} &=& - \varepsilon_{L}
\varepsilon_{S1}^{*} w_{1} - \imath \alpha
\varepsilon_{L} \varepsilon_{L}^{*} q_{2} \nonumber \\
& &- \imath \varepsilon_{L} \varepsilon_{S2}^{*} q_{3} - \imath
\varepsilon_{S2} \varepsilon_{S1}^{*} q_{2} \label{eq:scaled-q1},
\\
\frac{\partial}{\partial \tau} q_{2} &=& -  \varepsilon_{L}
\varepsilon_{S2}^{*} w_{2} + \imath \alpha
\varepsilon_{L} \varepsilon_{L}^{*} q_{1} \nonumber \\
& &- \imath  \varepsilon_{L} \varepsilon_{S1}^{*} q_{3}^{*} +
\imath  \varepsilon_{S1} \varepsilon_{S2}^{*} q_{1}
\label{eq:scaled-q2},
\\
\frac{\partial}{\partial \tau} q_{3} &=& \imath \varepsilon_{S2}
\varepsilon_{L}^{*} q_{1} + \imath
\varepsilon_{L} \varepsilon_{S1}^{*} q_{2}^{*} \nonumber \\
& & -  \varepsilon_{L} \varepsilon_{L}^{*}
(w_{1} - w_{2}) \nonumber \\
& & -  \varepsilon_{S2} \varepsilon_{S1}^{*} (w_{1} - w_{2})
\label{eq:scaled-q3}.
\end{eqnarray}
In these equations, the $q_{i}$'s are the molecular coherences for
the Raman-active transitions, the $w_{i}$'s represent the
population inversion in each of the two modes, and $\alpha$ is a
constant that describes the relative strength of the additional
Raman coupling. Specifically, we include the quasi-impulsive
pump-pump coupling, where the relative phase, $\phi_{L}$, appears
implicitly in the terms containing $\varepsilon_{L}
\varepsilon_{L}^{*}$. Finally, although substantial energy
conversion between pump and Stokes fields is possible, the
populations of the excited states in our experiments are always
much less than the ground state, so terms in the calculation
involving the off-diagonal coupling between the two excited states
($q_{3}$) may be suppressed.

In order to compare our model to the experimental results, we
numerically integrate the set of coupled differential equations
that describe the propagation of the fields inside the medium (see
Eqs. \ref{eq:scaled-full-pump} through \ref{eq:scaled-q3}).  We
set the amplitudes of the initial coherences ($q_{1}$ and $q_{2}$)
to be small random numbers, which in turn determine the initial
values of the Stokes phases $\phi_{S1}$ and $\phi_{S2}$. The
equations are integrated over an interaction region comparable to
the experimental conditions. At the end of the interaction length
we calculate the total energy contained in each of the Raman
fields as we vary the phase offset $\phi_{L}$ in our model pump
pulse.

We expect the initial phases of the Stokes fields to be random
because they buildup from spontaneous scattering. The stimulated
Stokes field emerging from the methanol cell is clearly multimode,
based on its angular divergence and far field appearance. When the
excitation volume corresponds to a large value of the Fresnel
number ($F = A / \lambda_{S} L$), such multiple spatial modes are
excited, leading to a large number of independent SRS processes in
each laser shot\cite{raymer:jul1985,scalora:feb1992}.

The high number of spatial modes implies multiple sets of initial
phases for the two different Stokes frequencies on every laser
pulse. Therefore we perform the integration over many trials of
random initial amplitudes of the coherences $q_{1}$ and $q_{2}$
(using a white Gaussian noise about zero). The control comes from
the $\varepsilon_{L} \varepsilon_{L}^{*}$ terms in the equations
for the temporal derivatives of $q_{i}$. The physical picture is
that the pump pulse drives the coherences $q_{i}$, which in turn
drive the Stokes modes $\varepsilon_{S1}$ and $\varepsilon_{S2}$.
Changing the value of $\phi_{L}$ in the equations changes which
mode receives a positive contribution and which receives a
negative. When the phase offset $\phi_{L}$ goes through $\pi/2$,
the role of the two modes switches.

\begin{figure}[ht]
\includegraphics [width=8.5cm, height=8.5cm, angle=0] {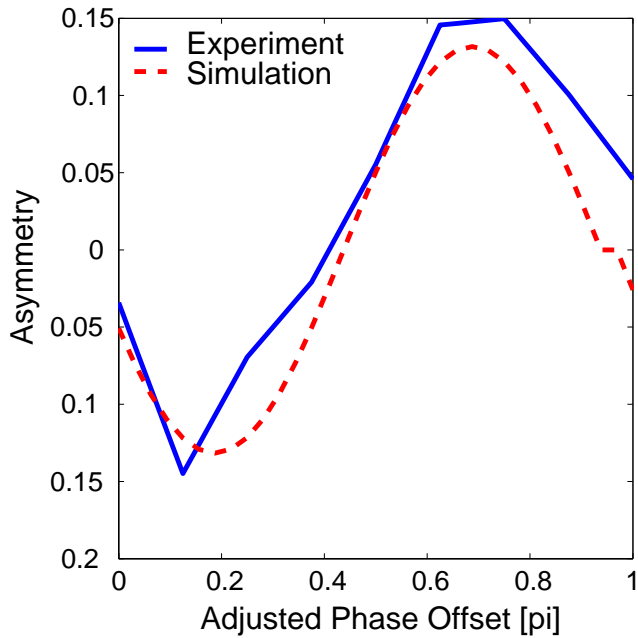}
\caption{Variation in mode excitation as a function of phase shift
between the two spectral regions for both experiment (solid) and
simulation (dashed). The vertical axis plots the difference in the
mode excitation divided by the sum.  The simulation results have
been adjusted to the data using two parameters: the initial phase
of the field and the coupling strength between the two excited
states.} \label{stokes:output}
\end{figure}

We define the mode asymmetry as the ratio of the difference in
energy between the two modes divided by the sum of the energy in
the two modes. Although for a given initial pair ($q_{1}$,$q_{2}$)
the mode asymmetry can take on any value between $\pm 1$, on
average one of the modes is preferentially excited at each value
of $\phi_{L}$. Figure \ref{stokes:output} shows the effect of mode
switching for the results from both the experiment and simulation.
The mode asymmetry is plotted along the vertical axis. We find
that both the experiment and simulation show a variation in mode
excitation with similar periods. The simulation results have been
adjusted to the data using two parameters: the pump-pump Raman
coupling strength ($\alpha$) and the initial phase offset
$\phi_{L}$. The depth of modulation in the asymmetry increases as
the Raman coupling strength increases (the plot shows the result
with $\alpha = 7$). Changing the phase offset $\phi_{L}$ simply
shifts the phase of the modulation. The phase offset in the
experiment is unknown due to pulse propagation effects within the
liquid. The mode control was relatively insensitive to the other
parameters of the model, which were set to approximate
experimental values of the Raman threshold.

In conclusion, we have demonstrated mode selection in stimulated
Raman scattering in liquid methanol through control of the
coherence of the pump laser.  Control is possible even though the
pump laser bandwidth is insufficient to seed the Stokes waves.
Using our feedback control results as a starting point, we have
constructed a simple model of the process that could explain the
effect. The control in our model comes from the Raman coupling of
the two excited modes by the pump laser.

We would like to acknowledge Chitra Rangan, Paul Berman, and Bruce
Shore for helpful discussions regarding the simulations. This work
is supported by the National Science Foundation under Grant No.
$9987916$.

\bibliography{brettbib}

\end{document}